\documentclass[aps,prd,preprintnumbers,nofootinbib,superscriptaddress,preprint]{revtex4-1}
\usepackage{graphicx}
\usepackage{amssymb}
\usepackage{amsmath}
\usepackage{color}
\usepackage{booktabs}
\usepackage{dcolumn}

\graphicspath{{fig/}}

\newcommand{\beq}{\begin{eqnarray}}
\newcommand{\eeq}{\end{eqnarray}}

\def\fsl#1{\setbox0=\hbox{$#1$}
   \dimen0=\wd0
   \setbox1=\hbox{/} \dimen1=\wd1
   \ifdim\dimen0>\dimen1
      \rlap{\hbox to \dimen0{\hfil/\hfil}}
      #1
   \else
      \rlap{\hbox to \dimen1{\hfil$#1$\hfil}}
      /
   \fi}

\begin{document}

\preprint{UTHEP-704, UTCCS-P-104}

\title{Relation between scattering amplitude and 
Bethe-Salpeter wave function in quantum field theory}

\date{\today
}

\author{Takeshi~Yamazaki}
\affiliation{Faculty of Pure and Applied Sciences, University of Tsukuba, \\ Tsukuba, Ibaraki 305-8571, Japan}
\affiliation{Center for Computational Sciences, University of Tsukuba, \\ Tsukuba, Ibaraki 305-8577, Japan}
\affiliation{RIKEN Advanced Institute for Computational Science,
Kobe, Hyogo 650-0047, Japan}
\author{Yoshinobu Kuramashi}
\affiliation{Center for Computational Sciences, University of Tsukuba, \\ Tsukuba, Ibaraki 305-8577, Japan}
\affiliation{RIKEN Advanced Institute for Computational Science,
Kobe, Hyogo 650-0047, Japan}

\begin{abstract}
We reexamine the relations between the Bethe-Salpeter (BS) wave function
of two particles, the on-shell scattering amplitude, and the
effective potential in quantum filed theory.
It is emphasized that there is an exact relation between 
the BS wave function inside the interaction range and the scattering amplitude,
and the reduced BS wave function, which is defined in this article,
plays an essential role in this relation.
Based on the exact relation, we show that 
the solution of Schr\"odinger equation with the effective potential
gives us a correct on-shell scattering amplitude 
only at the momentum where the effective potential is calculated,
while wrong results are obtained from the
Schr\"odinger equation at general momenta.
We also discuss about a momentum expansion of the reduced BS wave function
and an uncertainty of the scattering amplitude stemming from
the choice of the interpolating operator in the BS wave function.
The theoretical conclusion obtained in this article 
could give hints to understand
the inconsistency observed in lattice QCD 
calculation of the two-nucleon channels with different approaches.

\end{abstract}

\maketitle

\section{Introduction}

The finite volume (FV) method derived by 
L\"uscher~\cite{Luscher:1986pf,Luscher:1990ux}
is the most reliable method to evaluate the scattering phase shift
$\delta(k)$ in two-hadron system
from lattice QCD calculations on finite volume.
In this method $\delta(k)$ can be determined
from the relative momentum of two particles $k^2$ on finite volume.
A formula connecting $\delta(k)$ to $k^2$ on finite volume
was derived from the discussion of two-particle wave function
in the outside region of the interaction range in 
quantum mechanics~\cite{Luscher:1990ux}.
It was also confirmed that the same formula is derived 
from the Bethe-Salpeter (BS) wave function in 
quantum field theory~\cite{Lin:2001ek,Aoki:2005uf}.
Thus, the relation among 
$k^2$, $\delta(k)$, and the BS wave function in the outside region 
of the interaction range are well understood in the FV method.

On the other hand,
the HALQCD method~\cite{Aoki:2009ji}, which was recently proposed,
is another method to evaluate $\delta(k)$ from finite volume calculation
to utilize information of 
the BS wave function inside the interaction range.
In this method an effective potential defined by the BS wave function
is regarded as a potential in quantum mechanics.
The scattering phase shift $\delta(k)$ is obtained by solving the
Schr\"odinger equation in the infinite volume 
using the effective potential as an input.

In lattice QCD calculation of the two-nucleon channels,
two methods, the direct calculation of bound state energy 
based on the FV method and the HALQCD method, give qualitatively 
different results:
Bound states are confirmed by several independent groups with the former, while
no bound state is observed with the latter. 
See more details in recent reviews of the lattice conference~\cite{Walker-Loud:2014iea,Yamazaki:2015nka,Savage:2016egr}.
The reason of the inconsistency has not been understood at present,
though several possible reasons are suggested.

In order to understand the inconsistency,
we reexamine the relations between the following three quantities
in the framework of quantum field theory:
the BS wave function inside the interaction range,
the on-shell scattering amplitude, and the effective potential.
To do this, we first present the exact relation 
between the BS wave function inside the interaction range
and the on-shell scattering amplitude in quantum field theory,
which will be called the fundamental relation,
although it was already implicitly used in Ref.~\cite{Aoki:2005uf}.
It is emphasized that
the reduced BS wave function defined in Sec.~\ref{sec:def} 
plays an essential role in a direct relation to the on-shell 
scattering amplitude.
This fact means that
it is not necessary to introduce the effective potential
in quantum field theory for calculation of the scattering phase shift, 
even when we use the BS wave function inside the interaction range.
Based on this relation, we show that the
correct scattering phase shift is obtained from the HALQCD method 
only at the momentum
where the effective potential is defined,
while they are incorrect at other momenta.
Furthermore, we discuss about a momentum expansion 
of the reduced BS wave function
and an uncertainty of the scattering amplitude arising from 
the choice of operators for the BS wave function.
These discussions would be also helpful to understand what is obtained 
from the HALQCD method.

This article is organized as follows.
Section~\ref{sec:def} is devoted to give definitions for
various quantities in the framework of quantum field theory:
the BS wave function, the on-shell scattering amplitude,
the reduced BS wave function, and
the fundamental relation between them.
We clarify the relation between the two scattering phase shifts obtained
from the fundamental relation and the HALQCD method in 
Sec.~\ref{sec:schrodinger_eq}.
In Sec.~\ref{sec:expansion_h} we consider the expansion of the effective
potential and discuss the validity of the velocity expansion used in the HALQCD method.
In Sec.~\ref{sec:smearing_phi} an uncertainty of the scattering amplitude
coming from the choice of the operators in the BS wave function is discussed employing a smeared operator as an example.
We show that 
the smeared BS wave function gives a different scattering
amplitude from the one of the unsmeared BS wave function.
Section~\ref{sec:conclusions} summarizes conclusions.

\section{Definitions}
\label{sec:def}

\subsection{BS wave function and scattering amplitude}

In this article, we consider only the simplest two-particle
scattering, which is an $S$-wave scattering of 
two distinguishable scalar particles, whose mass is $m$, 
below the threshold $2E_p \le 4m$, where $E_p = \sqrt{m^2+p^2}$ with the relative momentum $p$.
We follow the definitions in Refs.~\cite{Lin:2001ek,Aoki:2005uf}.
The BS wave function 
in the infinite volume is defined by\footnote{Since we mainly discuss $\phi(\vec{x};\vec{k})$ in the infinite volume, the subscript $\infty$ of the BS wave function employed in Ref.~\cite{Aoki:2005uf} is omitted. On the other hand, we express the BS wave function on finite volume as $\phi_L(\vec{x};k)$.} 
\begin{equation}
\phi(\vec{x};\vec{k}) = \langle 0 | \pi_1(\vec{x}/2)\pi_2(-\vec{x}/2) |
\hat{\pi}_1(\vec{k})\hat{\pi}_2(-\vec{k}); {\rm in}\rangle,
\label{eq:defphixk}
\end{equation}
where $\pi_i$ is an interpolating operator of the $i$th scalar particle ($i=1,2$),
and $|\hat{\pi}_1(\vec{k})\hat{\pi}_2(-\vec{k}); {\rm in}\rangle$ is
an asymptotic two-particle state with the relative momenta $\vec{k}$
and $-\vec{k}$.\footnote{We omit $t$ dependence of $\phi(\vec{x};\vec{k})$, because it is expressed by an overall factor $e^{i 2E_k t}$. All the overall factors will be neglected in the later discussion.}

As discussed in Refs.~\cite{Lin:2001ek,Aoki:2005uf},
in quantum field theory,
the relation between $\phi(\vec{x};\vec{k})$ and 
the two-particle scattering amplitude $M(p;k)$ with $k = |\vec{k}|$
is derived using the LSZ reduction formula
in the case that inelastic scattering effects are negligible.
Here $M(p;k)$ is the off-shell amplitude defined by the 4-point
Green function:
\begin{eqnarray}
&&
e^{-i{\bf q}\cdot{\bf x}}
\frac{-i\sqrt{Z} M(p;k)}{-{\bf q}^2 + m^2 - i\varepsilon}=
\nonumber\\
&& \int d^4zd^4y_1d^4y_2 
K({\bf p},{\bf z}) K(-{\bf k_1},{\bf y_1}) K(-{\bf k_2},{\bf y_2})
\langle 0 | T[ \pi_1({\bf z})\pi_2({\bf x})\pi_1({\bf y_1})\pi_2({\bf y_2})
| 0 \rangle,
\label{eq:LSZ}
\end{eqnarray}
where the bold faced momenta and coordinates are four-dimensional vectors,
and 
\begin{equation}
K({\bf p},{\bf z}) = \frac{i}{\sqrt{Z}}e^{i{\bf p}\cdot{\bf z}}
(-{\bf p}^2 + m^2),
\end{equation}
with $Z$ the renormalization factor of the operator $\pi_i$.
As given in Ref.~\cite{Aoki:2005uf},
${\bf p}, {\bf k_1}$, and ${\bf k_2}$ are on-shell momenta,
\begin{equation}
{\bf p} = (E_p, \vec{p}),\ \ {\bf k_1} = (E_k, \vec{k}),\ \ 
{\bf k_2} = (E_k, -\vec{k}),
\end{equation}
while ${\bf q}$ is generally off-shell momentum and determined by 
the total energy momentum conservation,
\begin{equation}
{\bf q} = (2E_k - E_p, -\vec{p})
\end{equation}
with ${\bf q} = (E_p, -\vec{p})$ at on-shell.
$M(p;k)$ is related to the scattering
phase shift $\delta(k)$ at on-shell,
\begin{equation}
M(k;k) = \frac{16\pi E_k}{k}e^{i\delta(k)}\sin\delta(k) .
\label{eq:mkk}
\end{equation}
In Refs.~\cite{Lin:2001ek,Aoki:2005uf}
the relation between $\phi(\vec{x};\vec{k})$ and $M(p;k)$ is given by 
\begin{equation}
\phi(\vec{x};\vec{k}) = e^{i\vec{k}\cdot\vec{x}} +
\int\frac{d^3p}{(2\pi)^3}\frac{H(p;k)}{p^2-k^2-i\epsilon}
e^{i\vec{p}\cdot\vec{x}} ,
\label{eq:def_phixk}
\end{equation}
where $H(p;k)$ is related to $M(p;k)$ as
\begin{equation}
H(p;k) = \frac{E_p+E_k}{8E_pE_k}M(p;k).
\end{equation}
In Eq.~(\ref{eq:def_phixk}), irrelevant overall constants are neglected.
In this article we consider only the $S$-wave scattering so that
$\phi(\vec{x};\vec{k}) = \phi(x;k)$ where
the first term $e^{i\vec{k}\cdot\vec{x}}$ in the right-hand side 
of Eq.~(\ref{eq:def_phixk}) is replaced by 
the spherical Bessel function $j_0(kx)$.

\subsection{Reduced BS wave function}
\label{subsec:rBSwavefunction}

We define $h(x;k)$ using $\phi(x;k)$ as 
in Ref.~\cite{Aoki:2005uf}\footnote{$h(x;k)$ is defined in the opposite sign
to Eq.~(A10) in Ref.~\cite{Aoki:2005uf} to coincide the sign of 
lattice effective potential Eq.~(29) in the reference.},
\begin{equation}
h(x;k) = (\Delta + k^2) \phi(x;k) .
\label{eq:def_hxk}
\end{equation}
Using Eq.~(\ref{eq:def_phixk}) we also obtain
\begin{equation}
h(x;k) =
-\int\frac{d^3p}{(2\pi)^3} H(p;k)e^{i\vec{p}\cdot\vec{x}} .
\label{eq:hxk_Hpk}
\end{equation}
We assume that $h(x;k) = 0$ except for exponential tail 
in the outside region of the interaction range $R$
as in Ref.~\cite{Aoki:2005uf}.
We shall call $h(x;k)$ as
the reduced BS wave function in the following,
because $h(x;k)$ is more directly related to the scattering amplitude
defined through the reduction formula of Eq.~(\ref{eq:LSZ}) than
the BS wave function itself.
The reduced BS wave function plays an essential role
to calculate $\delta(k)$ in quantum field theory, when we 
use the information of $\phi(x;k)$ in $x \le R$,
which will be shown below.

\subsection{Fundamental relation}
\label{subsec:fundamentalrelation}

Using the definition of the reduced BS wave function Eq.~(\ref{eq:hxk_Hpk}),
$H(k;k)$, which is related to the on-shell amplitude Eq.~(\ref{eq:mkk}), 
is obtained
by the Fourier transformation of $h(x;k)$,
\begin{equation}
H(k;k) = -\int d^3x\, h(x;k) e^{-i\vec{k}\cdot\vec{x}} =
\frac{4\pi}{k}e^{i\delta(k)}\sin\delta(k) .
\label{eq:def_hkk}
\end{equation}
This is the relation between 
the reduced BS wave function $h(x;k)$, {\it i.e.}, 
$\phi(x;k)$ inside the interaction range, 
and the on-shell scattering amplitude $\delta(k)$ in
quantum field theory.
Although this relation
is not explicitly written in Ref.~\cite{Aoki:2005uf}, it was used to
show the relation between $\delta(k)$ and $\phi(\vec{x};\vec{k})$ in
$x \to \infty$ in the reference.
We shall call the relation as the fundamental relation through this article.
It should be noted that it is not necessary to introduce
the effective potential, which is defined in the next section,
for the calculation of $\delta(k)$ in quantum field theory,
if we use this relation using the reduced BS 
wave function $h(x;k)$.
 
In later sections, based on the fundamental relation, 
we will discuss a relation of $\delta(k)$ 
in quantum field theory and the one obtained from 
Schr\"odinger equation in quantum mechanics, and an uncertainty
arising from the choice of the interpolating operator in $\phi(x;k)$.

\section{Scattering phase shift from Schr\"odinger equation using effective potential}
\label{sec:schrodinger_eq}

\subsection{Effective potential}

The effective potential $V(x;k)$ was first introduced in 
Ref.~\cite{Aoki:2005uf}.
We follow its definition\footnote{The relation between $V(x;k)$ in this article and the effective potential $U(\vec{x};k)$ in Eq.~(29) of Ref.~\cite{Aoki:2005uf} is $V(x;k) = k^2 U(\vec{x};k)/m$.} with a factor $1/m$ assuming $V(x;k)=0$ in $x > R$,
\begin{equation}
V(x;k) = \left\{ 
\begin{array}{ll}
\displaystyle{\frac{1}{m}\frac{h(x;k)}{\phi(x;k)}} & (x \le R)\\
0 & (x > R)
\end{array}
\right. .
\end{equation}
Note that $V(x;k)$ diverges, if $\phi(x;k)$ has a node in $x \le R$.
Although we call it as the effective potential here, the relation between this quantity and the potential in quantum mechanics is not trivial; The former is the reduced BS wave function normalized by the BS wave function being manifestly momentum dependent, while the latter is defined to be momentum independent in principle. This is an essential difference between relativistic quantum field theory and nonrelativistic quantum mechanics\cite{Luscher:1986pf,Luscher:1990ux}.

In Ref.~\cite{Aoki:2005uf}
the BS wave function $\phi_L(\vec{x};k)$ in the $I=2$ two-pion channel
on finite volume was calculated in lattice QCD, and 
the effective potential was evaluated from $\phi_L(\vec{x};k)$.
In that study,
the effective potential was used only for the purpose 
to determine the interaction range $R$.
Using $\phi_L(\vec{x};k)$ outside the interaction range,
$k^2$ was obtained from two types of analyses:
$-\Delta \phi_L(\vec{x};k)/\phi_L(\vec{x};k)$ and
a fit of $\phi_L(\vec{x};k)$ with the form of the Green function 
on finite volume. Both gave consistent results and also agreed with the one from the two-pion energy.
Once $k^2$ on finite volume is determined,
$\delta(k)$ in the infinite volume is directly obtained through 
the FV method~\cite{Luscher:1986pf,Luscher:1990ux}.
The relations between $\delta(k)$, $k^2$ on finite volume, and $\phi_L(\vec{x};k)$
in $x > R$ are understood in
the FV method without any ambiguity.

The essential idea of the HALQCD method~\cite{Aoki:2009ji} is to regard $V(x;k)$ as a potential in quantum mechanics.
To obtain the scattering phase shift $\delta(p)$ at any $p$ below the threshold, which means that $p\ne k$ is available,
Schr\"odinger equation in the infinite volume is solved using $V(x;k)$ as 
a potential,
\begin{equation}
(\Delta + p^2) \overline{\phi}(x;p) = 2\mu V(x;k) \overline{\phi}(x;p),
\label{eq:schrodinger}
\end{equation}
where $\overline{\phi}(x;p)$ is the solution of the equation,
and $\mu$ is the reduced mass $\mu = m/2$.
$\delta(p)$ is obtained from the $x$ dependence of 
$\overline{\phi}(x;p)$ in $x \to \infty$.
Up to now, however, we have not known any explicit relation between the scattering phase shift $\delta(p)$ obtained from the Schr\"odinger equation with $V(x;k)$ and 
the one in the fundamental relation of Eq.~(\ref{eq:def_hkk}).
We will clarify this point in the following subsection.

\subsection{Scattering phase shift from Schr\"odinger equation}

As explained in a textbook of quantum mechanics\footnote{See for example Eq.~(7.1.34) in p.~384 of Ref.~\cite{Sakurai:1993zz}. Note that the normalization of the wave function in the equation differs by $(2\pi)^{3/2}$ from our case.},
if $V(x;k)$ in the Schr\"odinger equation of Eq.~(\ref{eq:schrodinger}) 
is a central potential going to zero faster than $1/x$ as $x\to\infty$,
the scattering amplitude $f(p)$ of the two-particle scattering with
the interaction $V(x;k)$ is given by 
\begin{equation}
f(p) = -\frac{2\mu}{4\pi}\int d^3x\, V(x;k) \overline{\phi}(x;p) 
e^{-i\vec{p}\cdot\vec{x}}=
-\frac{1}{4\pi} \int d^3x \frac{h(x;k)}{\phi(x;k)} \overline{\phi}(x;p) 
e^{-i\vec{p}\cdot\vec{x}}.
\label{eq:fp_phi}
\end{equation}
On the other hand, $f(p)$ can be also 
expressed by the scattering phase shift $\overline{\delta}(p)$,
\begin{equation}
f(p) = \frac{e^{i\overline{\delta}(p)}\sin\overline{\delta}(p)}{p} .
\label{eq:fp_delta}
\end{equation}

In the case of $p=k$, at which the effective potential $V(x;k)$ is defined, 
the Schr\"odinger equation of Eq.~(\ref{eq:schrodinger}) should
reduce to Eq.~(\ref{eq:def_hxk}) 
so that $\overline{\phi}(x;k) = \phi(x;k)$ holds in the Schr\"odinger equation
as well as in Eq.~(\ref{eq:fp_phi}).
Thus, $f(k)$ is written by $H(k;k)$ through 
the fundamental relation of Eq.~(\ref{eq:def_hkk}):
\begin{equation}
f(k) = -\frac{1}{4\pi} \int d^3x\, h(x;k)e^{-i\vec{k}\cdot\vec{x}} = 
\frac{1}{4\pi}H(k;k) =
\frac{e^{i\delta(k)}\sin\delta(k)}{k} .
\end{equation}
Comparing this equation with Eq.~(\ref{eq:fp_delta}), 
one obtains $\delta(k) = \overline{\delta}(k)$.
Therefore, the same scattering phase shift is obtained from the
Schr\"odinger equation and the fundamental relation of Eq.~(\ref{eq:def_hkk}) 
at the momentum $k$ where $V(x;k)$ is defined.

In the case of $p \ne k$, however, the two scattering phase
shifts obtained from the
Schr\"odinger equation with $V(x;k)$ and the fundamental relation
do not coincide anymore.\footnote{While in a special system, where $V(x;k)$ is assumed to be
independent of $k$ as defined in quantum mechanics, a correct $\delta(p)$ is obtained
through the Schr\"odinger equation,
in a general relativistic system one needs $V(x;p)$ at $p \ne k$
to obtain $\delta(p)$.}

\section{Expansion of reduced BS wave function}
\label{sec:expansion_h}

In quantum field theory, when we focus on $\phi(x;k)$ in $x \le R$,
most important quantity to calculate the on-shell scattering amplitude 
is the reduced BS wave function $h(x;k)$, 
because one can evaluate $\delta(k)$ from $h(x;k)$ 
through the fundamental relation of Eq.~(\ref{eq:def_hkk}).
A simple momentum expansion of $h(x;k)$ is allowed by the Taylor expansion 
in the vicinity of a given momentum.

On the other hand,
$h(x;k)$ is expressed by the nonlocal effective potential 
$U(x,x^\prime)$~\cite{Luscher:1990ux} as
\begin{equation}
h(x;k) = \int d^3x^\prime \, U(x,x^\prime) \phi(x^\prime;k) ,
\label{eq:def_u}
\end{equation}
where $U(x,x^\prime)$ does not explicitly depend on $k$~\cite{Aoki:2009ji}.
As in Ref.~\cite{Aoki:2009ji}, 
one can construct $U(x,x^\prime)$ by $k$ integration of Eq.~(\ref{eq:def_u}) 
up to the threshold momentum $k_{\rm max}$
with an appropriate function $g(x;k)$,
\begin{equation}
U(x,x^\prime) = \int_0^{k_{\rm max}} \frac{d^3 k}{(2\pi)^3} h(x;k) g(x^\prime;k),
\label{eq:u_integ_k}
\end{equation}
where $g(x;k)$ is assumed to satisfy
\begin{equation}
\int_0^{k_{\rm max}} \frac{d^3 k}{(2\pi)^3} 
\phi(x;k)g(x^\prime;k) = \delta(\vec{x}-\vec{x}^{\, \prime}).
\end{equation}
In the expression of Eq.~(\ref{eq:u_integ_k}),
the degree of freedom of $k$ in $h(x;k)$ is integrated out, and then it
is converted to the degree of freedom of $x^\prime$ in $U(x,x^\prime)$.

In the HALQCD method~\cite{Aoki:2009ji} $U(x,x^\prime)$ is approximated by
the so-called velocity expansion defined by
\begin{equation}
U(x,x^\prime) = \delta^3(\vec{x}-\vec{x}^{\, \prime}) 
(V_0(x) + V_1(x) \Delta + \cdots) ,
\label{eq:velocity_exp}
\end{equation}
where the dots express higher order terms of $O(\Delta^n)$ ($n\ge 2$),
and $V_i(x)$ is assumed to be independent of $k$.
This assumption provides a theoretical base for the time-dependent HALQCD method~\cite{HALQCD:2012aa},
where the $k$ independent $V_i(x)$ is extracted from $\phi(x;k)$
at any $k$ below the threshold.

We point out that the expansion parameter in this approximation is not clear. 
Formally, it is not an expansion in terms of the velocity, but of the Laplacian.
Substituting Eq.~(\ref{eq:velocity_exp}) into Eq.~(\ref{eq:def_u})
one obtains 
\begin{eqnarray}
h(x;k)&=&V_0(x)\phi(x;k) + V_1(x)(h(x;k)-k^2\phi(x;k)) + \cdots.
\end{eqnarray}
Since $h(x;k)$ appears both sides in the equation,
this is not a systematic expansion of $h(x;k)$ in terms of $k^2$
in contrast to the Taylor expansion. 
From the above equation we obtain the following expressions for the reduced BS wave function,
\begin{eqnarray}
h(x;k)&=&\left\{\frac{V_0(x)}{1-V_1(x)}-k^2\frac{V_1(x)}{1-V_1(x)}+O(k^0,k^2,k^4,\dots)\right\}\phi(x;k)\label{eq:h-expansion}\\
&=&\left\{V_0(x)+V_1(x)\Delta +O(k^0,k^2,k^4,\dots)\right\}\phi(x;k)\label{eq:h-expansion_x}\\
&=&[U(x,x^\prime)]^{\rm local}\phi(x;k). 
\end{eqnarray}
The coefficients of $k^{2n}$ ($n\ge 0$) should vary according to the inclusion of the higher order terms of $O(\Delta^m)$ ($m> n$) in Eq.~(\ref{eq:velocity_exp}). 
We point out that $[U(x,x^\prime)]^{\rm local}$, which is regarded as a local approximation of $U(x,x^\prime)$ in Eq.~(\ref{eq:def_u}), has $k$ dependence.
Since $h(x;k)/\phi(x;k)$ has the two independent degrees of freedom of
$x$ and $k$, $[U(x,x^\prime)]^{\rm local}$ needs to keep the same number of 
the degrees of freedom after the degree of freedom of $x^\prime$ is 
integrated out.\footnote{It is inappropriate to consider only the leading term $V_0(x)$, because $U(x,x^\prime)$ is a nonlocal potential.} 

The problem in this expansion becomes manifest in the practical determination of $V_i(x)$. The simplest example is the determination of the leading term $V_0(x)$, which is approximated by $h(x;k)/\phi(x;k)$\cite{Aoki:2009ji}.
 We find that it should contain the contributions of $O(k^{2n})$ ($n\ge 0$) from the higher order terms of the velocity expansion in order to properly describe the $k$ dependence of $h(x;k)$.\footnote{As discussed in Sec.~\ref{subsec:fundamentalrelation}, $h(x;k)$ directly provides the scattering amplitude through the fundamental relation of Eq.~(\ref{eq:def_hkk}) so that it does not make sense to approximate $V_0(x)$ by $h(x;k)/\phi(x;k)$ and solve the Schr{\"o}dinger equation with $V_0(x)$.} 
The inclusion of the term of $V_1(x)$ does not change the argument.
Both $V_0(x)$ and $V_1(x)$ are determined in $k$ dependent way with the use of $\phi(x;k)$ and $\Delta\phi(x;k)$. Otherwise the constructed $[U(x,x^\prime)]^{\rm local}$ has the uncertainties of $O(k^{2n})$ ($n\ge 0$). 
This is a consequence of the fact that $[U(x,x^\prime)]^{\rm local}$ has the two independent degrees of freedom of $x$ and $k$, even if $U(x,x^\prime)$ is superficially expanded using $\Delta$ and $V_i(x)$ without any explicit $k$ dependence.

\section{Uncertainty of scattering amplitude from the choice of operator in $\phi(\vec{x};\vec{k})$}
\label{sec:smearing_phi}

In this section, let us consider the case of smeared operator as a choice of operator in $\phi(\vec{x};\vec{k})$.
We discuss the relation between the scattering amplitude
and the BS wave function with the use of a smeared operator 
through the fundamental relation, and show that
the scattering amplitude obtained from the BS wave function
depends on the details of the smeared operator.

At first 
suppose $\phi(x;k)$ is constructed by the local operators. 
For convenience in the following discussion,
we change the coordinates of the two-particle operator in 
$\phi(x;k)$ of Eq.~(\ref{eq:defphixk})
from $\pi_1(\vec{x}/2)\pi_2(-\vec{x}/2)$ 
to $\pi_1(\vec{x})\pi_2(\vec{0})$.
This transformation does not loose any generality, because $\phi(x;k)$
is a function of the relative position of the two interpolating 
operators.
Thus, the modified $\phi(x;k)$ also satisfies Eq.~(\ref{eq:def_hxk}) yielding
the correct scattering amplitude in the fundamental relation.

For simplicity we will discuss the smeared-local BS wave function given by
\begin{equation}
\tilde{\phi}(\vec{x};\vec{k}) = \langle 0 |
\tilde{\pi}_1(\vec{x})\pi_2(\vec{0}) | \hat{\pi}_1(\vec{k})\hat{\pi}_2(-\vec{k});{\rm in} \rangle,
\end{equation}
where only one of the interpolating operators is smeared. The smeared operator
$\tilde{\pi}_1(\vec{x})$ is defined by
\begin{equation}
\tilde{\pi}_1(\vec{x}) = \int d^3y\, s(|\vec{x}-\vec{y}|) \pi_1(\vec{y})
\end{equation}
with $s(x)$ a smearing function as $s(x)\to 0$ for $x \to \infty$.
As discussed in Sec.~\ref{sec:def}
we consider only the $S$-wave scattering so that 
$\tilde{\phi}(\vec{x};\vec{k}) = \tilde{\phi}(x;k)$.
The Laplacian of $\tilde{\pi}(\vec{x})$ gives
\begin{equation}
\Delta \tilde{\pi}_1(\vec{x}) = \Delta_x \int d^3y\, s(|\vec{x}-\vec{y}|) \pi_1(\vec{y})
= \int d^3y\, \left[\Delta_y s(|\vec{x}-\vec{y}|)\right] \pi_1(\vec{y})
= \int d^3y\, s(|\vec{x}-\vec{y}|) \left[ \Delta_y \pi_1(\vec{y}) \right],
\end{equation}
where $\Delta_x$ is the Laplacian with respect to $\vec{x}$, and
we use $\Delta_x s(|\vec{x}-\vec{y}|) = \Delta_y s(|\vec{x}-\vec{y}|)$ and
the integration by parts in the second and the third equalities, respectively.
The corresponding reduced BS wave function $\tilde{h}(x;k)$ is defined as
\begin{equation}
\tilde{h}(x;k) = (\Delta + k^2) \tilde{\phi}(x;k) = 
\int d^3y\, s(|\vec{x}-\vec{y}|) h(y;k).
\end{equation}
The Fourier transformation of $\tilde{h}(x;k)$ gives 
a different scattering amplitude from that for $h(x;k)$ in the fundamental relation of Eq.~(\ref{eq:def_hkk})
\begin{equation}
\tilde{H}(k;k) = -\int d^3x\, \tilde{h}(x;k) e^{-i\vec{k}\cdot\vec{x}}
= -\int d^3x\int d^3y\, s(|\vec{x}-\vec{y}|) h(y;k) e^{-i\vec{k}\cdot\vec{x}},
\label{eq:def_tildeh_kk}
\end{equation}
which explicitly depends on $s(x)$.\footnote{It is pointed out in Ref.~\cite{Kawai:2017goq} that the $s(x)$ dependence can be factorized by using the Fourier transformation of $s(x)$, {\it i.e.}, $\tilde{H}(k;k) = C(k) H(k;k)$, where $C(k) = \int d^3x\, s(x) e^{-i\vec{k}\cdot \vec{x}}$.}

Assuming that $s(x)$ is not much broad and there is a region where 
$\tilde{h}(x;k) = 0$ in $x > \tilde{R}$,
the effective potential $\tilde{V}(x;k)$ is given by
\begin{equation}
\tilde{V}(x;k) = \frac{1}{m}\frac{\tilde{h}(x;k)}{\tilde{\phi}(x;k)}
\end{equation}
in $x \le \tilde{R}$ and $\tilde{V}(x;k) = 0$ in $x > \tilde{R}$.
Following the discussion in Sec.~\ref{sec:schrodinger_eq},
the Schr\"odinger equation with $\tilde{V}(x;k)$
gives the amplitude $\tilde{H}(k;k)$ of Eq.~(\ref{eq:def_tildeh_kk}) 
at the momentum $k$.
The amplitude $H(k;k)$ cannot be obtained from $\tilde{V}(x;k)$, when we use Eq.~(\ref{eq:fp_phi}).

Therefore, a smearing of the interpolating operator in the BS wave 
function gives
a different scattering amplitude from the one obtained from 
the fundamental relation, 
which depends on the smearing function $s(x)$.
One can easily expect that the smeared-smeared BS wave function,
where the two interpolating operators are smeared,
also gives a different scattering amplitude from $H(k;k)$ as well as 
$\tilde{H}(k;k)$, and
its relation to $h(x;k)$ becomes more complicated 
than Eq.~(\ref{eq:def_tildeh_kk}).
Furthermore, a smearing of the quark field in the interpolating operator
of hadrons utilized in lattice QCD calculations
make the relation between the BS wave function and 
the scattering amplitude far more complicated.

The above discussion leads to the conclusion 
that the smearing of the interpolating 
operators in the BS wave function causes
uncertainties both in the use of 
Eq.~(\ref{eq:fp_phi}) and the fundamental relation.
On the other hand, in the FV method, 
the relevant quantity is essentially $k^2$ on finite volume, 
which can be obtained 
even with the use of the smeared operators.  For example
we can extract $k^2$ from
$-\Delta \tilde{\phi}(x;k)/\tilde{\phi}(x;k)$ in $x > \tilde{R}$
as shown in Ref.~\cite{Aoki:2005uf}.

\section{Conclusion}
\label{sec:conclusions}

We have reexamined the relations between the BS wave function
inside the interaction range, the on-shell scattering amplitude, and
the effective potential in the framework of quantum field theory.
The fundamental relation allows us to obtain the scattering phase shift
$\delta(k)$ from the reduced BS wave function $h(x;k)$ which
is essentially the BS wave function inside the interaction range,
while we obtain $\delta(k)$ by analyzing the BS wave function 
outside the interaction range in the FV method.

The fundamental relation tells us two important facts.
First, it is not necessary to introduce the effective potential
$V(x;k)$ for calculation of $\delta(k)$ in quantum field theory.
In case that one uses $\phi(x;k)$ in $x\le R$ to extract $\delta(k)$, 
most relevant quantity is $h(x;k)$.
Second, the solution at the momentum $p$ of the Schr\"odinger equation with
$V(x;k)$ as a potential gives the correct scattering phase shift 
only at $p = k$, while it is incorrect at $p \ne k$.
In other words, the correct $\delta(k)$ is obtained only from $V(x;k)$ 
defined at the same momentum as $\delta(k)$,
and it is not allowed to use $V(x;k)$ to calculate $\delta(p)$ at $p \ne k$.

It is possible to expand $h(x;k)$ by the Taylor expansion in
the vicinity of a given momentum.
On the other hand, the velocity expansion is not a systematic 
expansion of $h(x;k)$ in terms of $k^2$. 
The coefficients of the velocity expansion $V_i(x)$ should depend on $k$
to correctly describe the $k$ dependence of $h(x;k)$.

We have also shown that 
the BS wave function using a smeared operator yields
a modified scattering amplitude, which depends on the smearing function.
This conclusion holds true for both methods: the direct use of 
the fundamental relation and the solution of the Schr\"odinger equation with the effective potential.

The results obtained in this article are also 
applicable to the bound state, and
can be extended to systems of particles with 
nonzero spin~\cite{Ishizuka:2009bx}.
Our theoretical conclusion could be helpful to understand the inconsistency 
observed in the lattice QCD results of the two-nucleon channels
obtained by the two methods: the direct calculation based on the FV method
and the HALQCD method.

\section*{Acknowledgements}
T.Y. thanks Naruhito Ishizuka for fruitful discussions.
This work is supported in part by Japan Society for the Promotion of Science (JSPS) Grants-in-Aid for Scientific Research for Young Scientists (A) No. 16H06002.

\bibliographystyle{apsrev4-1}
\bibliography{reference}

\end{document}